\documentclass[twocolumn,showpacs,preprintnumbers,amsmath,amssymb,superscriptaddress,10pt,aps,prb]{revtex4-2}
\usepackage{setspace}
\usepackage{epsfig,amsopn}
\usepackage{graphicx}
\usepackage{epstopdf}
\usepackage{sidecap}
\usepackage{amsmath,amssymb}
\usepackage{amsthm}
\usepackage{enumerate}
\usepackage{xcolor}
\usepackage[colorlinks=true,linktoc=page,linkcolor=blue,citecolor=blue,urlcolor=blue,]{hyperref}
\usepackage{mathtools}
\usepackage[compact]{titlesec}
\usepackage{fancyhdr}
\titlespacing{\section}{20pt}{5ex}{2ex}
\begin{document}
	\title{\Large{Structural-modulation-driven Curie-temperature enhancement in Cr-doped SrRuO$_3$}}
	\author{Pooja}
	\email{kpooja@iitk.ac.in}
	\affiliation{Department of Physics, Indian Institute of Technology Kanpur, Kanpur 208016, India}
	\author{Bikash Saha}
	\affiliation{Solid State Physics Division, Bhabha Atomic Research Centre, Mumbai, 400085 India.}
	\author{Nitesh Choudhary}
	\affiliation{Department of Polymer and Process Engineering, Indian Institute of Technology Roorkee, Saharanpur Campus, 247001, U.P., India}
	\author{Pradip K.Maji}
\affiliation{Department of Polymer and Process Engineering, Indian Institute of Technology Roorkee, Saharanpur Campus, 247001, U.P., India}
	\author{A. K. Bera}
	\affiliation{Solid State Physics Division, Bhabha Atomic Research Centre, Mumbai, 400085 India.}
	\author{S. M. Yusuf}
	\affiliation{Solid State Physics Division, Bhabha Atomic Research Centre, Mumbai, 400085 India.}
		\author{Chanchal Sow}
	\email{chancahal@iitk.ac.in}
	\affiliation{Department of Physics, Indian Institute of Technology Kanpur, Kanpur 208016, India}
    \date{\today}
\begin{abstract}
		 Strongly correlated system with competing ground states are often poised close to the quantum critical point. External perturbations such as pressure, strain, electric field, and chemical doping can stabilise its ground state with exotic physical properties. Cr-doping is the lone exception which enhances the Curie-temperature in one of such correlated system SrRuO$_3$. To find the origin of $T_C$ enhancement, we investigate temperature-dependent structure, spectroscopic, magnetic and magnetotransport properties in SrRu$_{1-x}$Cr$_x$O$_3$. Cr-doping squeezes the unit cell volume which effectively enhances the stretching octahedral distortion by nearly five times than pure SrRuO$_3$. The Curie temperature increment by $\sim$ 22 K for x = 0.15 is found to be intertwined with the structural-modulation. Temperature-dependent Neutron diffraction analysis indicate that the unit cell volume minima coincide exactly with the enhanced ferromagnetic ordering ($\sim$ 190 K). Further analysis reveals that the effect of Cr-doping not only freezes the octahedral tilt below 100 K but also suppresses the complex magnetism responsible for exchange bias and topological hall effect in SrRuO$_3$. The spectroscopic measurements find a reduction of itinerancy of d-electrons with Cr-doping. The magnetotransport measurements portray an evolution from itinerant to localised ferromagnetism.  
	\end{abstract}
	\maketitle
	\section{Introduction}
\begin{figure}[t]
\centering
\includegraphics[width=0.379\linewidth]{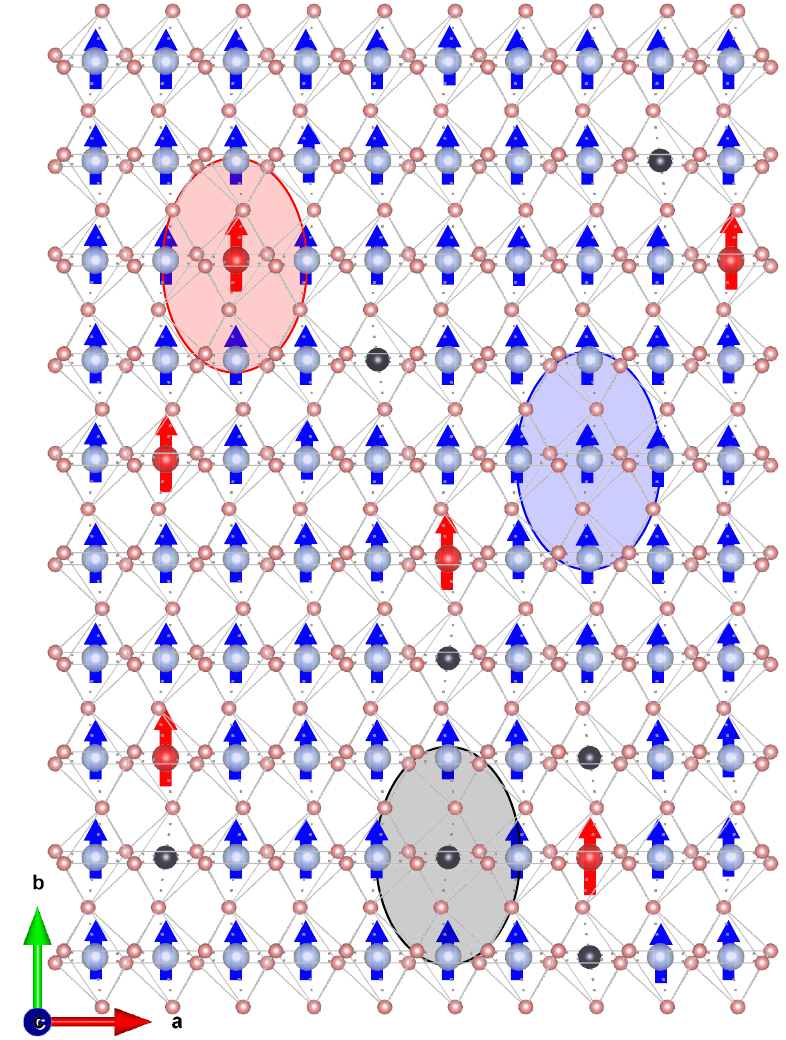} 
 \caption{(color online) Schematic representation of 10$\times$10$\times$1 cell of SrRu$_{1-x}$Cr$_x$O$_3$. Here only the Ru (light blue balls) and oxygen (light red balls) atoms are shown for clarity. The Cr atoms  (shown in red and black balls) are randomly distributed in the Ru matrix. The magnetic moments of Ru$^{4+}$ (S=1) and Cr$^{3+}$ (S=3/2) are shown with blue and red arrows respectively, and are ferromagnetically aligned along the b-axis. Cr$^{6+}$ (S=0) are shown as black balls that are non-magnetic and act as a disorder. In such a system, there are two dominant exchanges: the blue encircled region has pure Ru-Ru FM exchange, and the Red region has Ru-Cr FM exchange. The grey region (contains the non-magnetic Cr) is mainly involved in reducing the range of exchange interaction whereas the red region enhances the strength of the exchange interaction than SRO.}
 \label{fig:1}
\end{figure} 
Quantum materials with tunable multi-functionalities is an attractive area of research due to its intriguing magnetic, electronic and optical characteristics \cite{valenzuela2012novel,liu2013origin,salama1989high}. Such diverse physical properties bring vast possibilities for applications in microelectronic, magnetic and spintronic devices \cite{rothberg2011integrated}. Tuning the relative strength of spin-orbit coupling ($\lambda$) and on-site Coulomb interaction ($U$) from low to high, various quantum phases have emerged \cite{witczak2014correlated}. In particular, ruthenates with moderate strength of $\lambda$/$U$ become interesting candidates as they lie near the crossover region. The layered Ruthenates, especially in Ruddlesden-Popper (RP) series: (Sr/Ca)$_{1+n}$Ru$_n$O$_{3n+1}$, where $n$ (1 $<$ $n$ $<$ $\infty$) is the number of Ru-O layers, exhibit a drastic change of physical properties with a change in $n$. In the Ca-based RP series, a robust antiferromagnetic (AFM) insulating state (Mott Insulator \cite{nakatsuji19972}) is stabilized for $n$ = 1, which gradually decreases with increasing $n$. Thus the magnetic state of metallic CaRuO$_3$ (end member) is critically poised among AFM, ferromagnetic (FM) and paramagnetic (PM) \cite{longo1968magnetic,kiyama1998specific}. Sr$_2$RuO$_4$ ($n$ = 1) is a nonmagnetic metal down to 2 K and becomes a $p$-type superconductor below 1.5 K \cite{cao1997observation}. For $n$ $>$ 2, the ferromagnetic metallic state is stabilized in Sr-based RP series and the curie temperature is enhanced with $n$. The end member SrRuO$_3$ ($n$ = $\infty$) is an itinerant ferromagnet below $T_C$ $\sim$ 166 K \cite{longo1968magnetic}. SrRuO$_3$ identifies with distorted orthorhombic ($Pnma$) perovskite structure with sizeable $U$ (= 3-3.5 eV \cite{dang2015electronic}) and $\lambda$ (= 40-80 meV \cite{annett2006spin}). In addition, SrRuO$_3$ also possesses strong hybridization between Ru-4$d$ and O-2$p$ orbitals. These features facilitate a diverse range of intriguing physical characteristics in SrRuO$_3$  \cite{laad2001origin,wang2004anomalous,palai2009observation,kanbayasi1976magnetic,reich1999spin}. 
The recent finding of topological Hall effect and magnetic skyrmions in SrRuO$_3$ thin-films arising due to the interfacial strain has gained considerable interest \cite{qin2019emergence,matsuno2016interface,lu2021defect,wang2018ferroelectrically}.  
\par In bulk systems the external perturbations such as pressure, temperature, internal strain by chemical doping, and electric field change the physical properties drastically. For example, the crystal structure undergoes perovskite ($Pbnm$) to post-perovskite ($Cmcm$) structure accompanied by a 1.9 \% volume collapse with application of high pressure ( $\sim$ 40 GPa ) \cite{zhernenkov2013pressure}, which also alters the magnetism by decreasing the Ru-O bond lengths \cite{pietosa2008pressure}. On the other hand, temperature also drives two structural phase transitions from orthorhombic to tetragonal (at 800 K) and from tetragonal to cubic (at 975 K) \cite{chakoumakos1997high}. Chemical doping provides an additional method for tuning the magnetic exchange by internally deforming the structure. The size of the cation can be chosen higher or lower than the Ru depending on the requirement of unit cell volume expansion or contraction \cite{shepard1997thermodynamic}. This also affects the orthorhombicity and distortion in RuO$_6$ octahedra thereby a significant change in the magnetism \cite{cao1997itinerant,he2001disorder,petr2010physical}. In particular SrRu$_{1-x}$M$_x$O$_3$, where M (= Zn$^{2+}$, Mn$^{3+}$, Co$^{2+}$, Ni$^{2+}$ and Ti$^{4+}$) is transition metal ions (TMI) suppresses the $T_C$ down to 100 K \cite{pi2002substitution, gupta2017site}. Doping Cr is an exceptional case (among TMI) that enhances the $T_C$ of SrRuO$_3$ up to $\sim$ 190 K \cite{han2005nuclear, cao1996itinerant}. The enhancement of $T_C$ ($\sim$ 210 K) is also noticed in Pb-doped SrRuO$_3$, the origin is found to be structural distortion \cite{cao1996itinerant}. However, the origin of increased $T_C$ with Cr-doping in SrRuO$_3$ is still not clear. A schematic picture of magnetic moments with possible exchange interactions in SrRu$_{1-x}$Cr$_x$O$_3$ is depicted in Fig-\ref{fig:1}. Experimental reports suggest the mixed valence state of ruthenium, which creates Ru$^{4+/5+}$-O-Cr$^{4+/3+}$ minority double exchange interaction and enhances the ordering temperature \cite{han2005nuclear,dabrowski2005increase,zhang2011role}.
 While the First-principles calculation (using LSDA, LSDA+U and GGA+U) explain the origin of the increased T$_C$ in SrRu$_{1-x}$Cr$_x$O$_3$ system in terms of Cr$^{4+}$ - Ru$^{4+}$ hybridization \cite{hadipour2010effect} and p-d coupling \cite{wang2010first}. \par In order to find a clear picture of such T$_C$ enhancement, we present a detailed and systematic structural, spectroscopic, magnetic, transport, and microscopic magnetic state by Neutron diffraction studies on single phase SrRu$_{1-x}$Cr$_x$O$_3$ (0 $<$ x $<$ 0.15) samples. Our study established a strong connection between structural modulation with the increased $T_C$ in Cr-doped SrRuO$_3$.  
\section{Sample preparation and experimental details-} High quality, SrRu$_{1-x}$Cr$_x$O$_3$ (0 $<$ x $<$ 0.15) (SRCO) samples are synthesized via solid-state reaction method. Powders of SrCO$_3$ ($>$ 99.95$\%$), RuO$_2$ ($>$ 99.9$\%$), and Cr$_2$O$_3$ ($>$ 99.9\%) are used as ingredients. The mixture is pelletized and sintered for 48 hours at 900 - 1400 $^{\circ}$C with multiple intermediate grindings. RuO$_2$ is volatile as it sublimates at 1200$^{\circ}$C. To compensate, the extra 5 weight \% of RuO$_2$ is added to preserve the stoichiometric ratio. The atomic percentage of Ru and Cr is examined by the electron probe microscopy analyzer (EPMA) equipped with JXA-8230; and JEOL (as tabulated in Table- \ref{t:1}). The Phase purity and crystallinity of all samples are checked by using a high-resolution X-ray powder diffractometer (PAN analytical Empyrean Cu K$\alpha$).
\begin{table}[b]
\centering
\caption{EPMA results: Atomic percentage of Ru\% and Cr\% with the nominal ( x$_{nom}$) and determined (x$_{det}$) value of x in SrRu$_{1-x}$Cr$_x$O$_3$ samples.}
\begin{tabular}{|c|c|c|c|c|}\hline
Sample & x$_{nom}$&x$_{det}$ &Ru\% & Cr\%   \\ 
\hline
SRO &0 &0 & 102.0&0   \\
SRCO5&0.05&0.07&87.9&7.06\\
 SRCO10&0.10&0.11&90.32&11.51\\
SRCO15 &0.15  & 0.15&86.52& 15.86 \\
\hline 
\end{tabular}
\label{t:1}
\end{table}    
The valence states of Cr, Ru, and Sr are verified by PHI Versa Probe-II X-ray photoelectron spectroscopy (XPS) consisting of a monochromatic Al K$\alpha$ (h$\nu$ = 1486.6 eV) source. The magnetic and transport measurements are conducted between 2-300 K temperature and 0-6 T of applied magnetic field range using commercial quantum design MPMS (MPMS-XL) and PPMS respectively. Temperature-dependent (5-300 K) Neutron diffraction (ND) measurements are carried out on a high-quality SRCO15 polycrystalline sample using a powder neutron diffractometer, PD-I, at Dhruva Research Reactor, Mumbai (India). The powder sample is filled in a cylindrical vanadium can and attached to the cold finger of the He-4 gas-based closed-cycle refrigerator. 
\section{Calculation details-} The spin polarised density of states (DOSs) calculation is carried out on SrRuO$_3$ by using generalised gradient approximation (GGA) and GGA+U method with \href{https://www.quantum-espresso.org/}{Quantum Espresso} \cite{giannozzi2020quantum}. The Hubbard U term which is the Coulomb repulsion between on-site electrons, is considered to understand the effect of electron correlation on DOSs. We considered the conventional cell (Z = 4) of SrRuO$_3$ consisting of 20 atoms with lattice parameters obtained from Rietveld-refined XRD-data for SRO and SRCO15 (without any structural relaxation calculation). The calculation is performed with 4$\times$4$\times$4 K-points, by using  Perdew–Burke–Ernzerhof (PBE) exchange–correlation functional \cite{perdew1996generalized}. The ultra soft pseudo potentials (USPP), and a basis set for valence electrons consisting of plane waves with cut-off energy 25 Rydberg (340 eV) are used for self-consistent calculation. To understand the electron correlation effect on DOS the value of $U$ is taken 3.5 eV \cite{wang2010first} for Ru atom. 
\section{Results and discussion-}
\subsection{Structural analysis-}   
\begin{figure*}
\begin{center}
\includegraphics[width=0.89\linewidth]{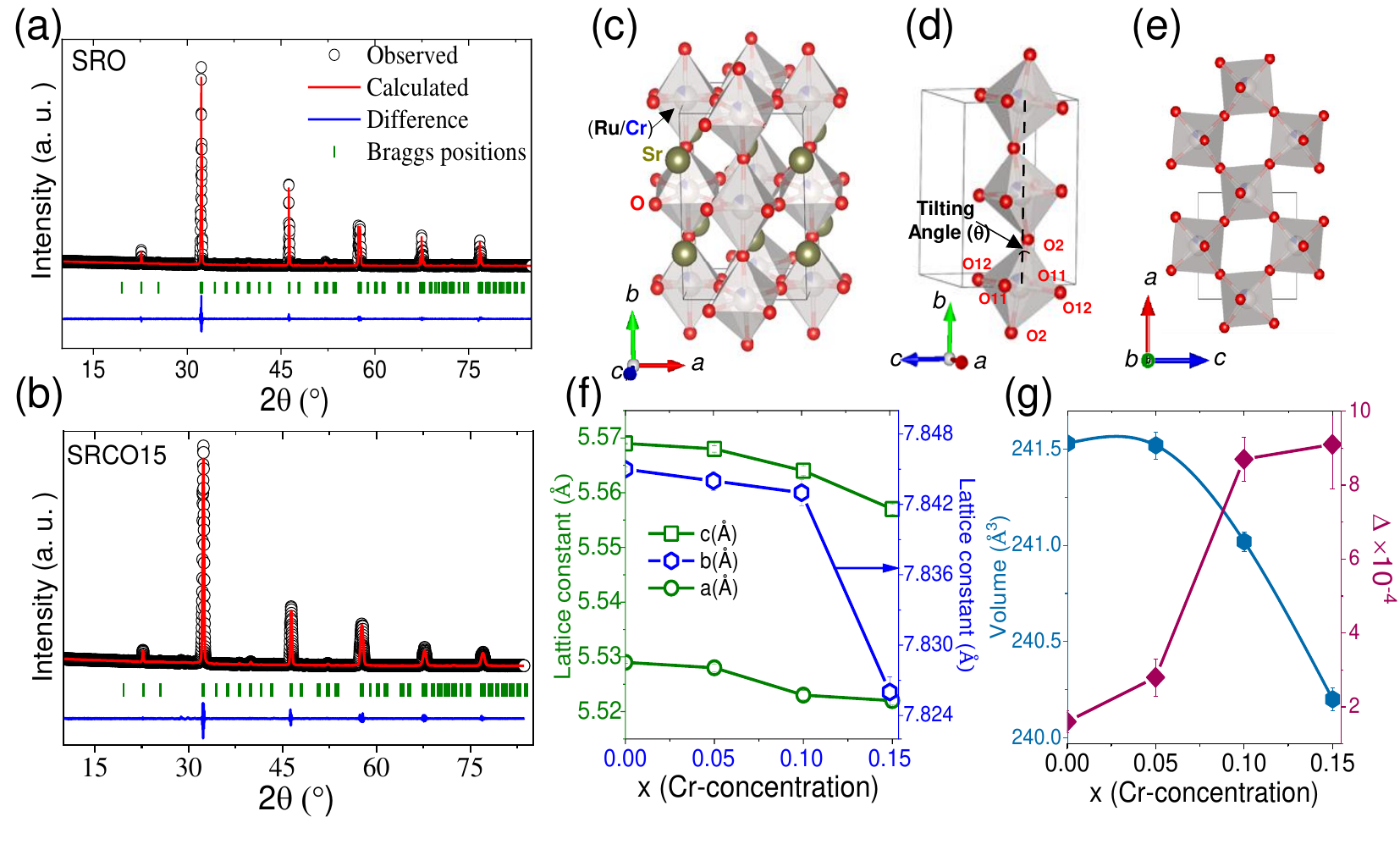} 
  \caption{(color online) Rietveld refined X-Ray diffraction pattern of (a) SRO and (b) SRCO15. The open black circles and red solid lines are observed and calculated patterns respectively and the blue line shows the difference between observed and calculated patterns, whereas green vertical bars are allowed Bragg peak positions. (c) The orthorhombic crystal structure of SRCO15 along with (Ru/Cr)O$_6$ octahedral arrangement. The crystal structure is drawn using \href{https://jp-minerals.org/vesta/en/}{VESTA} visualizing program \cite{momma2008vesta}, from the output file after rietveld refinements. Schematics of  (Ru/Cr)O$_6$ octahedra to demonstrate the octahedral tilt and various oxygen atoms in the (d) $bc$-plane and (e) $ac$-plane. The black solid line represents the unit cell. Variation of (f) lattice constants, (g) unit cell volume (left y-axis) and $\Delta$ (right-y axis) is shown upon Cr doping.}
 \label{fig:N1}
\end{center}
\end{figure*} 
Figure-\ref{fig:N1}(a) and (b) displays the experimentally measured and calculated Rietveld-refined (using \href{http://www.ill.eu/sites/fullprof/}{FULLPROF} \cite{ritter2011neutrons}) X-ray diffraction patterns of the SRO and SRCO15 respectively. Rietveld analysis reveals that SRCO samples have an orthorhombic crystal symmetry ($a$ $\neq$ $b$ $\neq$ $c$, $\alpha$ = $\beta$ = $\gamma$ = 90) with space group $Pnma$ (62) \cite{dabrowski2005increase,williams2006charge}. The Cr atom occupies the same Wyckoff-4$a$ site as the Ru atom. The schematic crystal structure is shown in Fig-\ref{fig:N1}(c). The crystal structure consists of corner-sharing identical (Ru/Cr)O$_6$ octahedra with tilt along all three directions as shown in Fig-\ref{fig:N1}(d, e). Fig-\ref{fig:N1}(f) shows the variation of lattice constant with respect to Cr-content. At 15\% Cr-doping the lattice constants $a$, $b$, and $c$ decrease by $\sim$ 0.13\%, 0.24\%, and 0.22\% respectively. As a result, the overall $\sim$ 0.54\% decrease in unit cell volume is shown in Fig-\ref{fig:N1}(g). The decrease in the unit cell volume is attributed to the smaller cationic radius of Cr$^{6+}$(0.044 nm)or Cr$^{3+}$(0.061 nm) (confirmed by XPS) compared to Ru$^{4+}$(0.062 nm). The refined crystal structure reveals three distinct octahedral bond lengths namely Ru-O11 (basal), Ru-O12 (basal), and  Ru-O2 (apical) with values 1.976, 2.0596 $\AA$ and 1.929 respectively in case of SRCO15. The unequal bond lengths give rise to the octahedral distortion \cite{lu2013role}. The octahedral stretching distortions ($\Delta$) and angular distortions ($\Sigma$) can be calculated using the equations:
\begin{equation}
\Delta = \frac{1}{6}\Sigma^6_{i=1}
\Big[\frac{|d_i - d_{mean}|}{d_{mean}}\Big]^2
\label{eq:1}
\end{equation}
\begin{equation}
\Sigma = \frac{1}{12} \Sigma^{12}_{i=1}|\phi_i - 90|
\label{eq:2}
\end{equation}
respectively, where $d_{mean}$ and $d_i$ denote the average and $i^{th}$ Ru/Cr-O bond length (there is a total of 6 different Ru-O lengths in a RuO$_6$ octahedra) respectively. The angle $\phi_i$ is considered as $i^{th}$ O-(Ru/Cr)-O bond angles in the (Ru/Cr)O$_6$ octahedra. The stretching distortion increases as we increase the Cr-doping in SRO. At 15\% Cr-doping, the value of $\Sigma$ is increased by 22\%, whereas $\Delta$ is enhanced by nearly 5 times compared to parent compound SRO. Such increased distortion due to Cr-doping affects the superexchange/double-exchange interaction between Ru moments. In this regard, theoretical work by $\emph{L. Wang et al.}$ suggests for low Cr-doped samples (x$<$ 0.25), Cr$^{3+}$ couples ferromagnetically with Ru$^{4+}$ moments which is responsible for enhancement of the $T_C$ \cite{wang2010first}. The calculated value of $\Sigma$, $\Delta$ and derived lattice parameters of SRCO15 in comparison with SRO at room temperature are tabulated in Table-\ref{t:2}.
\begin{table*}
\centering
 \caption{Rietveld-refinement structural parameters of SRO and SRCO15 in comparison with the room-temperature ND results.}
\begin{tabular}{|c|c|c|c|c|c|c|c|c|}\hline
  Sample&a&b&c&V&Ru-O1-Ru&Ru-O2-Ru&$\Sigma$&$\Delta$\\
  &($\AA$)&($\AA$)&($\AA$)&($\AA^3$)&($^{\circ}$)&($^{\circ}$)& &\\
  \hline
  SRO&5.529&7.845&5.569&241.5&163.9&159.1&4.5&1.6$\times 10^{-4}$\\
  SRCO15&5.522&7.826&5.557&240.2&163.5&157.9&5.5&9.1$\times 10^{-4}$\\
  \hline
\end{tabular}
\label{t:2}
\end{table*} 
 \subsection{X-ray photoelectrons spectroscopy-}
\begin{table*}
\centering
 \caption{The Binding energy (eV) values of different constituent elements in SRO and SRCO15.}
\begin{tabular}{|c|c|c|c|c|c|c|c|}\hline
  Sample&Ru3$d_{5/2}$&Ru3$d_{3/2}$&Sr3$p_{1/2}$&Cr2$p_{3/2}$&Cr2$p_{1/2}$&Sr3$d_{5/2}$&Sr3$d_{3/2}$\\
 &(eV)&(eV)&(eV)&(eV)&(eV)&(eV)&(eV)\\
 \hline
 SRO&280.75&285.95&278.0&-&-&132.21&135.21\\ 
  &282.05&288.55& -&-&-&133.76&136.34\\ 
\hline 
  SRCO15&281.35&286.05&278.75&577.86 (Cr$^{3+}$&587.92(Cr$^{3+}$)&132.39&134.94\\
  &282.6&288.5&-&581.54(Cr$^{6+}$)&590.69(Cr$^{6+}$)&133.64&136.34\\
  \hline
\end{tabular}
\label{t:3}
\end{table*}   

\begin{figure}
\begin{flushleft}
\includegraphics[width=\linewidth]{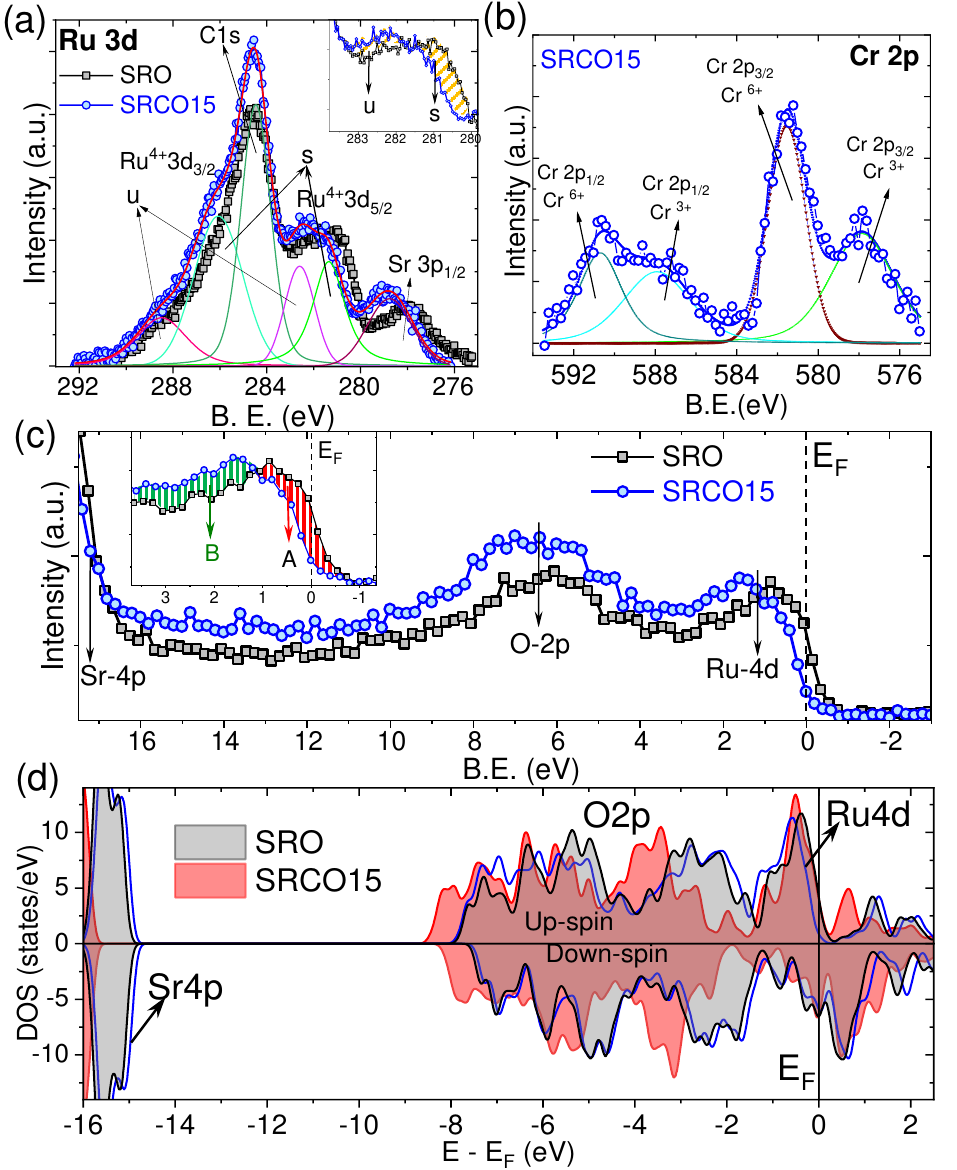} 
\caption{(color online) Core level XPS-spectrum of (a) Ru-3d in SRO and SRCO15, (b) Cr-2p in SRCO15, where the open circle shows the raw data and the solid lines are peak sum, and different coloured peaks represent the fittings (using \href{https://xpspeak.software.informer.com/4.1/} {XPSPEAK-4.1} \cite{yang2006carbon} ). All fitting peaks are labelled according to their B.E. positions in all curves. (c) XPS valence band spectrum of SRO and SRCO15. The inset figure in a, and c represent the zoomed-in-view of spectra. (d) Theoretically calculated TDOS spectra for SRO with U = 0 (gray), U = 3.5 eV (blue-line), and SRCO15 with U = 0 (red)}
\label{fig:N2}
\end{flushleft}
\end{figure}

The X-ray photoelectron spectroscopy (XPS) is employed to investigate the valence states of various constituent elements and the density of state (DOS) in the vicinity of Fermi energy (E$_F$). The XPS survey scan identifies the presence of Sr, Ru, Cr, and O in SRCO samples (not shown here). Fig-\ref{fig:N2}(a) depicts the Ru-3$d$ XPS core spectra of SRO and SRCO15 (normalised the intensity by removing the Shirley background after fitting). The peak at $\sim$ 278 eV is agreed as Sr3$p_{1/2}$ and the other two peaks in the Ru-3d spectra corresponds to spin-orbit doublet (SOD): Ru 3$d_{5/2}$ and Ru 3$d_{3/2}$ with splitting energy $\Delta$E$_{SO}^{Ru}$ $\sim$ 5 eV, suggesting Ru$^{4+}$ valence state \cite{morgan2015resolving} in all samples. Each Ru doublet has two peaks labelled as screened ($s$) and unscreened ($u$). The $s$ peak is generally present at $\sim$ 2 eV lower binding energy (B.E.) than $u$ peak, in metallic systems due to the screening effect of core electrons \cite{kim2004core}. The presence of both peaks also indicates that the samples are metallic at room temperature. It is important to note that in the case of SRCO15 the Ru core spectra are shifted towards the higher binding energy. The shift is visible in the zoomed-in-view of Ru 3$d_{5/2}$ peak in the inset of Fig-\ref{fig:N2}(a). This spectral shift arises due to the reduction of delocalized electrons \cite{brar2023lattice} in the system with increasing disorder via Cr-doping. The Sr-3d spectra (not shown) have fitted with two SOD peaks Sr3d$_{5/2}$ and Sr3d$_{3/2}$ confirming the presence of Sr$^{2+}$ state. Fig-\ref{fig:N2}(b) shows the Cr-2p core spectra of SRCO15. The Cr SOD peaks labelled as Cr 2p$_{1/2}$ and Cr 2p$_{3/2}$ are separated by $\Delta$E$_{SO}^{Cr}$ $\sim$ 9.2 eV. These doublets are further split into two peaks corresponding to the two different oxidation states as Cr$^{3+} 3d^3_{t_{2g}}$ and Cr$^{6+} 3d^0_{t_{2g}}$ (splitting energy $\sim$ 3.68 eV). The peak intensity ratio of Cr$^{6+}$ and Cr$^{3+}$ peaks is found to be $\sim$ 2:1 (Fig-\ref{fig:N2}(b)). 
The overall valence state of Cr is estimated by the relation: Cr$^{6+}_{x-y}$Cr$^{3+}_y$ = Cr$^{4+}_x$. It maintains an overall 4+ oxidation state at the Ru site. For SRCO15, y $\sim$ 0.045, suggesting one-third of Cr is in the 3+ state and two-thirds is in the 6+ oxidation state. Thus overall charge neutrality equation for SRCO15: Sr$^{2+}$Ru$^{4+}_{0.85}$Cr$^{4.5+}_{0.15}$O$^{2-}_{3 + \delta}$, where $\delta$ $\sim$ 0.04 (calculated from average charge valency of Oxygen \cite{wu2018study}). The B.E. values of Ru, Cr, and Sr of SRO and SRCO15 are tabulated in Table-\ref{t:3}. Fig-\ref{fig:N2}(c) shows the comparative valence band spectra of SRO and SRCO15. The peak centred $\sim$ 1.2 eV and 6.2 eV correspond to the Ru-4d/Cr-3d and O-2p electrons respectively, indicating the strong hybridization between Ru/Cr and oxygen \cite{maiti2006role}. The inset Fig-\ref{fig:N2}(c) shows a zoomed-in-view of spectral region at $\sim$ 1.2 eV, which is marked as feature-A (red) and B (green) ascribed as coherent (delocalized) and incoherent (localized) electrons respectively. In SRCO15 the spectral weight of feature-A near $E_F$ is shifted to feature-B, suggesting the decrement of a delocalized electron with increasing Cr-doping in SRO. Such shift of the delocalized electron mainly arises due to electron correlation \cite{singh2007manifestation}. To understand the effect of electron correlation and structure distortion in the DOS with Cr-doping, we theoretically calculated the DOS. The total density of state (TDOS) for SrRuO$_3$ is calculated with $U$ = 0, $U$ = 3.5 eV and with modified lattice parameters at 15\% Cr-doping (with U = 0) as shown in Fig-\ref{fig:N2}(d).
The calculated TDOS shows half- metallic state and the bands between 2.5 eV to -9 eV energy containing Ru-4d, O-2p electronic contributions consisting with the earlier reported results \cite{wang2010first,hadipour2010effect}. The up-spin TDOS spectra resemble the experimentally obtained valence band spectra. It is crucial to note that in both cases: with finite $U$ and modified lattice parameter (for SRCO15), the delocalised electron near the $E_F$ is decreased and shifted towards the localised state. Thus the Cr-doping modifies the crystal structure and induces the electron correlation in the system hence reducing the number of itinerant electrons. This localised behaviour of charge carriers is further understood through the magnetic and transport properties of SRCO.
\begin{table*}
\centering
 \caption{ Magnetic parameters of SRO and SRCO15.}
\begin{tabular}{|c|c|c|c|c|c|c|c|c|}\hline
 Parametrs/& $M_{5K}$ &$T_C$&$H_c$&$M_s$&$A_{SW}$&$B_{SE}$&J$\times$k$_B$&$T_C^{cal}$\\
 Samples& (emu/g) &(K)&(kOe)&($\mu_B$/f.u.)&($K^{-3/2}$)&(K$^{-2}$)& &(K)\\
 \hline
  SRO &17.6&166.2&3.30&1.42&1.33$\times$10$^{-4}$&2.3$\times$10$^{-6}$&28.98&158.6\\
  SRCO15&13.4&188.5&2.88&1.05&1.22$\times$10$^{-4}$&2.04$\times$10$^{-6}$&30.7&167.9\\
  \hline
\end{tabular}
\label{t:4}
\end{table*}
\subsection{Magnetic measurements-} 
\begin{figure}
\begin{flushleft}
\includegraphics[width=\linewidth]{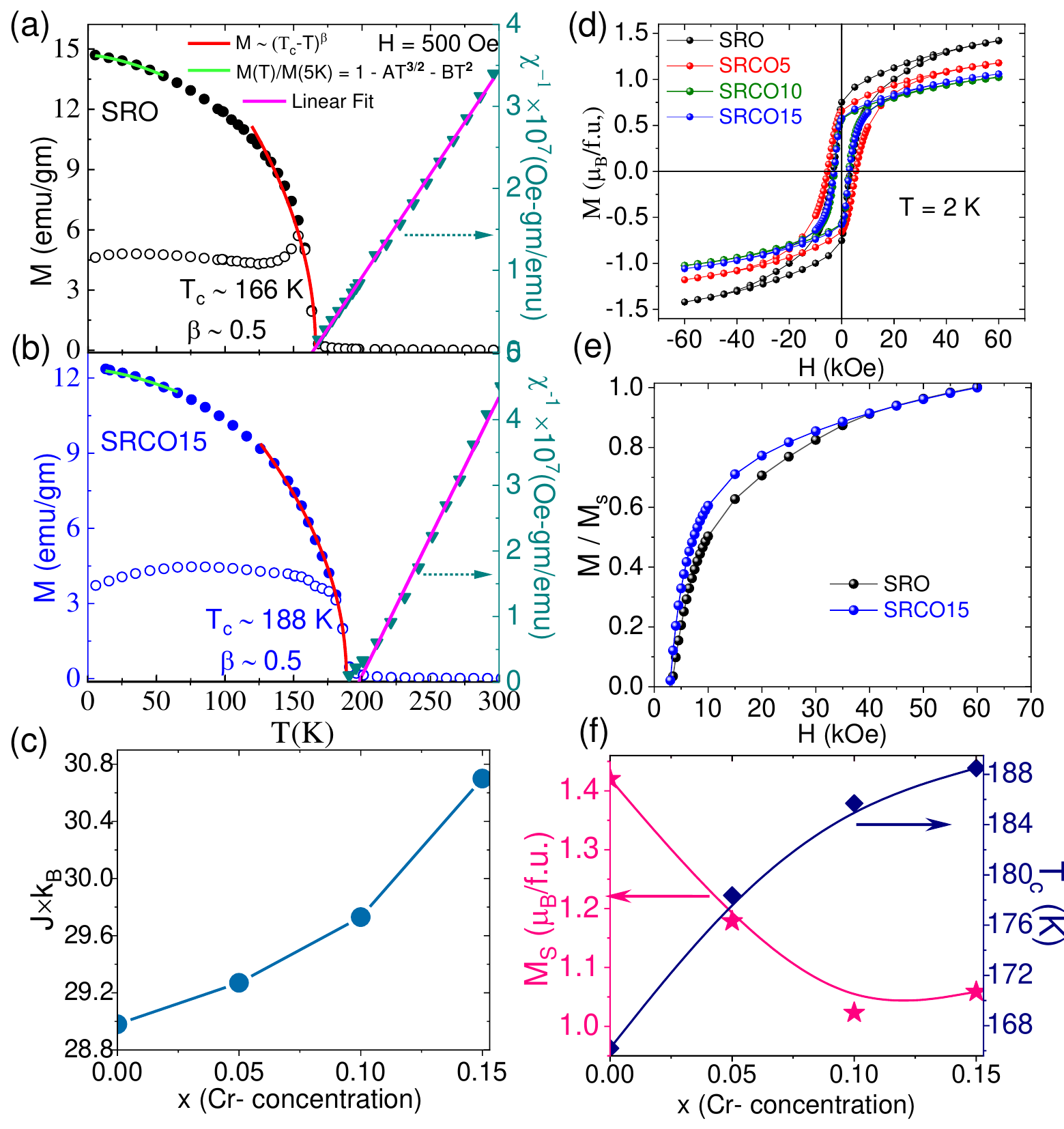} 
\caption{(color online) (Left y-axis) Temperature-dependent FC and ZFC magnetization of (a) SRO and (b) SRCO15 measured at 500 Oe, where solid red (green) lines are FC-magnetization fitting near critical region (below 50 K). (Right y-axis) inverse FC-susceptibility, where pink solid lines show the Curie-Weiss linear fit. (c) Variation of $J \times k_B$ with Cr-doping. (d) M-H hysteresis loops of SRCO samples at 2 K. (e) Comparative plot of M/M$_s$ vs field for SRO and SRCO15. (f) Variation of  T$_C$ (right y-axis) and M$_s$ (left y-axis) at 60 kOe with different Cr concentrations.}
 \label{fig:N3}
\end{flushleft}
\end{figure}

The structural distortion and the mixed valence state of Cr ion certainly change the magnetic exchange interaction in SRCO. Fig-\ref{fig:N3}(a) and (b) shows the temperature-dependent FC-ZFC magnetization data of SRO and SRCO15 respectively. Both exhibit FM behaviour. The FM Curie-temperature ($T_C$) is calculated by fitting the FC magnetisation data with the scaling law: $M$ $\approx$ ($T_C$ - $T$)$^{\beta}$, in the vicinity of the critical region. The value of $T_C$ turns out to be 166 K and 188 K for SRO and SRCO15 respectively. The critical exponent $\beta$ is $\sim$ 0.5 for both samples indicating mean field type FM-interactions. Most importantly 22 K enhancement in the $T_C$ is noticed, whereas the FC magnetization at 5 K ($M_{5K}$) is reduced with Cr-doping (see Table-\ref{t:4}). At low temperatures, the thermally excited magnons (following  Bloch $T^{3/2}$ law), as well as Stoner excitations (following $T^2$ law), reduce the magnetic moment. Hence the low temperature magnetization data fitted with: $M(H, T)$ = $M(H,0)$[1 - $A_{SW}$ $T^{3/2}$ - $B$ $_{SE} T^2$], where $A_{SW}$ is a spin-wave stiffness constant and $B_{SE}$ is Stoner excitation parameter. The value of $A_{SW}$ is related with the exchange constant ($J$) between two Ru$^{4+}$ atom as: $A_{SW}$ = (0.0587/S)($k_B/2JS$)$^{3/2}$ and with $T_C$ as: ($A_{SW}$ )$^{-2/3}$= 2.42 $T_C$ \cite{snyder2019critical}. The obtained fitting values of $A_{SW}$ and $B_{SE}$ decreases, $J$ and T$_C$ increases with increasing Cr-doping in SRO (see Table-\ref{t:4}). In Fig-\ref{fig:N3}(c) the exchange constant, $J \times k_B$ is plotted against the Cr-doping, which shows a gradual increase and explains the $T_C$ enhancement.\\
 Fig-\ref{fig:N3}(d) shows the M-H hysteresis plot at 2 K for SRCO samples. The magnetization increases up to a saturation magnetic moment with increasing the magnetic field. For SRO the magnetic moment at 6 T (M$_{s}$) is $\sim$ 1.42 $\mu_B/f.u.$. This ordered moment is lower than the anticipated complete S = 1 moment (2 $\mu_B$/Ru), which may arise due to several reasons: (a) itinerant ferromagnetic nature \cite{cao1997thermal}, (b) spin glass state at low-temperature \cite{reich1999spin}, and (c) large magnetocrystalline anisotropy (MCA). In case of SRCO15, the resultant magnetic moment is reduced by $\sim$25\% than the parent compound SRO. In SRCO15 the origin of the reduced moment can be understood with the presence of Cr$^{6+}$ (S = 0). It is important to note that the total spin-only moment is reduced (as 10\% spin is getting zero). This is in agreement with the fact that an FM-like hysteresis is observed with a non-saturating component for all SRCO samples. The non-saturating behaviour in SRO is supported by large MCA. In Cr-doped SRO, due to the randomness of exchange interactions: [(a) FM with Ru$^{4+}$-O-Ru$^{4+}$,  Ru$^{4+}$-O-Cr$^{3+}$ (b) PM Ru$^{4+}$-O-Cr$^{6+}$, and rare possibility of (c) AFM Cr$^{3+}$-O-Cr$^{3+}$] a non-saturation of the magnetisation is expected. Fig-\ref{fig:N3}(e) shows the comparative plot of normalized magnetization (M/M$_{s}$) as a function of field (0 - 6 T). It is evident from the plot that SRO possesses a higher MCA than SRCO15. The origin of higher anisotropy in SRO compared to SRCO15 is intriguing. Below 100 K various kinds of magnetic anomalies were observed in bulk as well as in thin film SrRuO$_3$. In bulk, SrRuO$_3$ the spin glass, memory effect, slow domain wall dynamics, anomalous magneto transport, and Hall effect sign inversion have been observed while in thin film exchange bias, Skyrmionic behaviour, topological Hall effect, etc. have been reported \cite{han2023reversal,qin2019emergence}. All these observations suggest the formation of a complex magnetic order below 100 K in SRO. The origin of this ``complex ferromagnetism" (CFM) was found to be srtuctural-modulation driven \cite{sow2012structural}. The effect of Cr-doping reduces the anisotropy field in SRO. Such reduction in the anisotropy field is related to the reduction of CFM in SRCO15, which is discussed in the magnetotransport section. Fig-\ref{fig:N3}(f) shows the variation of $M_{s}$ (at 2 K, 6 T) as well as $T_C$ as a function of Cr-concentration. The value of M$_{s}$ is decreasing with increasing Cr-doping in SRO, whereas $T_C$ is increasing. All obtained values of $T_C$, $A_{SW}$, $B_{SE}$,  $J$, and $M_{FC}$(5 K) of SRCO15 in comparison with SRO are tabulated in Table-\ref{t:4}. To understand the microscopic magnetism and evolution of magnetic structure the temperature-dependent ND measurement is performed. 
 \begin{figure*}
\begin{flushleft}
\includegraphics[width=\linewidth]{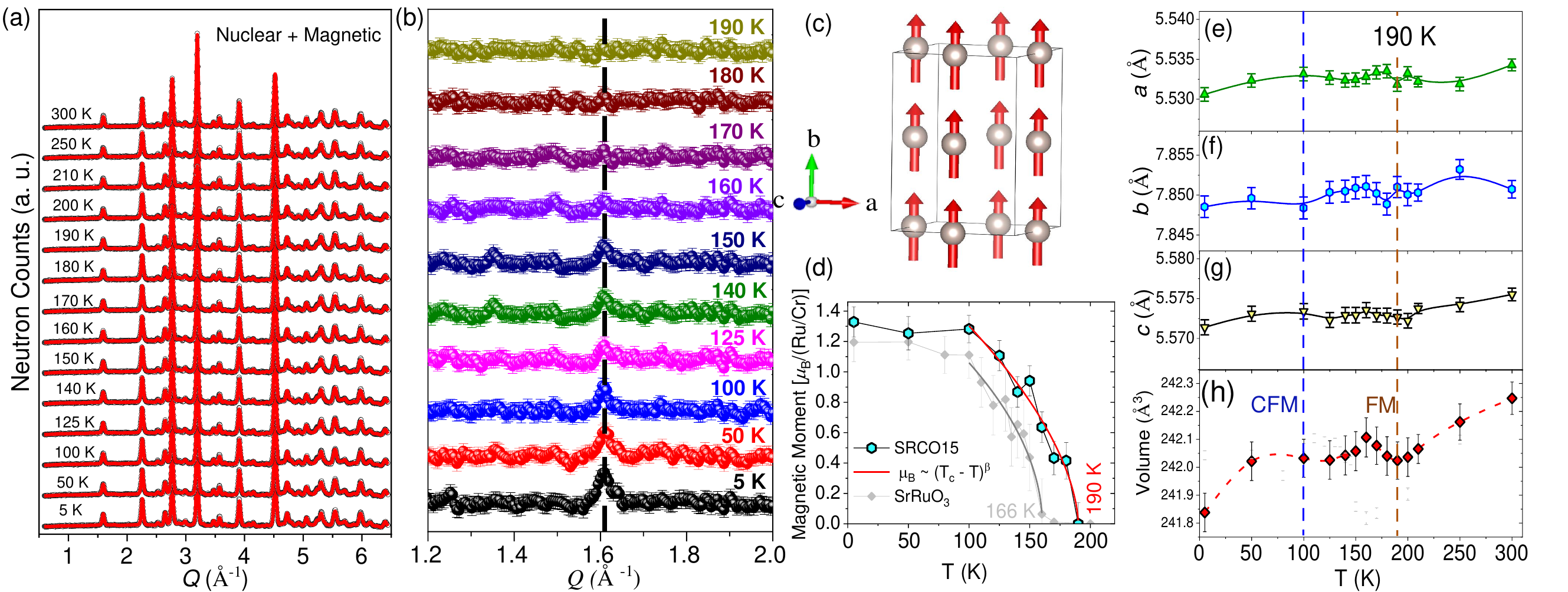} 
 \caption{(color online) (a) Neutron diffraction patterns (Nuclear + magnetic) recorded over 5 – 300 K temperature range. (b) Temperature evolution of magnetic Bragg peak at ~ 1.6 $\AA^{-1}$. (c) The magnetic structure of SRCO15. (d) Temperature dependence of the ordered magnetic moment of SRCO15 (blue hexagon data) at the Ru site. The dashed line is the magnetic exponent fitting. Here SrRuO$_3$ data (gray) \cite{sow2012structural}, is plotted for comparison. (e - h) Temperature evolution of the lattice constants, and unit cell volume of SRCO15, where colored vertical dotted lines at 190 K (orange) and 100 K (blue) represent onset temperature of ferromagnetic (FM) and complex ferromagnetic (CFM) ordering respectively.}
 \label{fig:N13a}
\end{flushleft}
\end{figure*}
\subsection{Neutron diffraction-} 
\begin{figure}
\begin{flushleft}
\includegraphics[width=\linewidth]{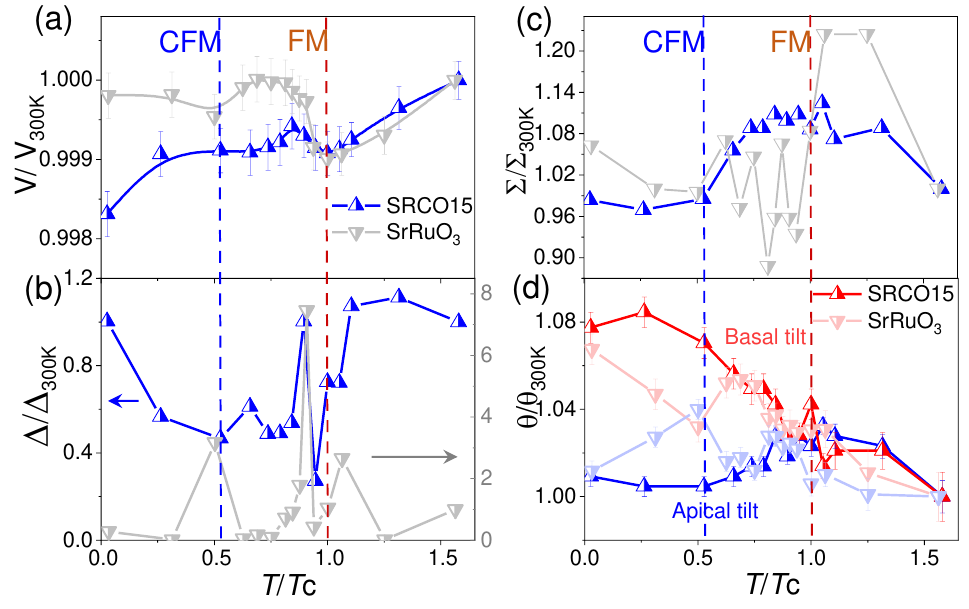} 
 \caption{(color online) The temperature variation of (a) unit cell volume and (b) stretching distortion, (c) angular distortion of Ru/CrO6 octahedra and (d) octahedral tilt angle ($\theta$) for SRCO15 and SRO compounds. The values are normalized with respect to their reference values at 300 and 250 K for SRCO15 and SRO, respectively.  The data for SRO are taken from \cite{sow2012structural, sow2013freezing}.}
\label{fig:N14}
\end{flushleft}
\end{figure}

Figure-\ref{fig:N13a}(a) shows the experimentally recorded and Rietveld-refined ND patterns for the SRCO15 sample at various
temperatures (5 – 300 K). The peaks at $\emph{Q}$ $\sim$ 1.6 $\AA^{-1}$ (2$\theta$ = 16$^{\circ}$) in the neutron diffraction data correspond to magnetic Bragg peaks (101) and (020). To get the pure magnetic signal the nuclear background (PM-state) taken at 250 K is subtracted from the low-temperature data (below the magnetic ordered state). Fig-\ref{fig:N13a}(b) shows such background subtracted data for various temperatures. The intensity of the magnetic peak ($\emph{Q}$ $\sim$ 1.6 $\AA^{-1}$) starts to appear below 190 K and attains a maximum at 5 K. This reveals a ferromagnetic ordering with a propagation vector, \textbf{$\emph{k}$} = (0,0,0), as reported for the parent compound SrRuO$_3$ \cite{sow2012structural}. To determine the symmetry of all the allowed magnetic structures of SRCO15, representation analysis \cite{bera2017zigzag,bera2022magnetism,bera2011effect} is carried out using the BASIREPS program \cite{ritter2011neutrons}. The symmetry analysis for the space group $Pnma$ and the propagation vector \textbf{$\emph{k}$} = (0,0,0) reveal four possible non-zero irreducible representations ($\Gamma_1$, $\Gamma_3$, $\Gamma_5$, and $\Gamma_7$) (Table-\ref{t:5}). Out of these four irreducible representations, the best agreement of the experimentally observed to the Rietveld refined magnetic diffraction pattern is achieved for $\Gamma_5$. Since, the experimental value of the magnetic form factor for Ru$^{4+}$ is not available in the literature, therefore the magnetic form factor of Ru$^{1+}$ is used to solve the magnetic refinement of the diffraction patterns of SRCO15 (similar to the magnetic refinement for the several other Ru$^{4+}$ based compounds \cite{ranjan2009magneto,bushmeleva2006evidence}). The magnetic structure obtained from the Rietveld analysis of the ND pattern is shown in Fig-\ref{fig:N13a}(c). From the magnetic structure, it is evident that the moments are lying predominantly along the $b$-axis. The component of the moments along the $\emph{a}$ and $\emph{c}$ maximize is found to be negligible ($<$ 0.1 $\mu_B$/Ru$^{4+}$). The refined ordered moment value at 5 K is found to be; $M$ = 1.328 $\pm$ 0.1 $\mu_B$/Ru$^{4+}$, closely matches with the experimental ( M$_s$ value of SRCO15 i.e. $\sim$ 1.1 $\mu_B$/Ru$^{4+}$). The temperature dependence of the ordered magnetic moment of SRCO15 and in comparison with SRO is shown in Fig-\ref{fig:N13a}(d). The fitting of the power law equation, $\mu_B$(T) = A(T$_C$-T)$^{\beta}$, finds $T_C$ = 190 K and  $\beta$ = 0.52 (closely agree with experimental value ($T_C$ = 188 K, and $\beta$ = 0.5). From ND it is evident that the Cr-doping results $\sim$ 22 K enhancement of FM ordering temperature, however, the mean field-like behaviour remains the same. It has to be noted that the ordered magnetic moment is considerably lower than the spin-only moment 2 $\mu_B$/Ru$^{4+}$. \\
\begin{table}[b]
\centering
 \caption{Basis vectors for the atoms of the magnetic sites 4b site of the orthorhombic space group $Pnma$, with the propagation vector, \textbf{$\emph{k}$} = (0,0,0).}
\begin{tabular}{|c|c|c|c|c|}\hline
 & $\psi_1$ & $\psi_2$ (-x+$\frac{1}{2}$, & $\psi_3$ (-x, & $\psi_4$ (x+$\frac{1}{2}$,\\
 & (x, y, z) & -y, z+$\frac{1}{2}$) & y+$\frac{1}{2}$, -z) & -y+$\frac{1}{2}$, -z+$\frac{1}{2}$)\\
 \hline
 $\Gamma_1$&(1, 1, 1)&(-1, -1, 1)&(-1, 1, -1)&(1, -1, -1)\\
 $\Gamma_3$&(1, 1, 1)&(-1, -1, 1)&(1, -1, 1)&(-1, 1, 1)\\
 $\Gamma_5$&(1, 1, 1)&(1, 1, -1)&(-1, 1, -1)&(-1, 1, 1)\\
 $\Gamma_7$&(1, 1, 1)&(1, 1, -1)&(1, -1, 1)&(1, -1, -1)\\
 \hline
\end{tabular}
\label{t:5}
\end{table} 
 Now begin with the question of what is the origin of enhancement in $T_C$. The answer lies within the evolution of structure parameters with temperature. The lattice parameters are obtained from the Rietveld analysis of all ND patterns. The variation of lattice parameters and the unit cell volume is plotted against temperature as shown in Fig-\ref{fig:N13a}(e - h). A gradual contraction of the lattice constants and unit cell volume has been observed with the decrease in the temperature. The volume contraction below the ordering temperature can be understood as a spontaneous magnetostriction effect that arises due to the magnetic ordering of the atomic moments below $T_C$. The unit cell volume resembles a minima at 190 K. A minima in unit cell volume represents a stable configuration (minimum free energy of the system), when a magnetic ordering is usually formed. Fig-\ref{fig:N14}(a) shows the comparative volume plot of SRCO15 and SrRuO$_3$. In the case of SRO, an anomaly in the lattice parameters and unit cell volume at the ferromagnetic ordering temperature (166 K). It is important to note that 0.5 \% volume contraction occurs only due to Cr doping and lowering the temperature further enhances the contraction affecting the $\Sigma$ and $\Delta$. Thus Cr doping results in a relative change in the crystal structure with respect to temperature. Such change is visible in the form of a minimum in volume and a peak in $\Delta$ near 190 K (22 K higher than the parent compound SRO). Thus the origin of the $T_C$ enhancement is closely associated with structural modulations due to both Cr-doping and temperature. Further, in the case of SrRuO$_3$, another minimum below 100 K ($\sim$ 75 K) is noticed, which is responsible for the CFM in addition to the dominant FM order. However, Cr-doping suppresses the formation of such CFM. This picture is more clear in the transport section later.
Further, the structural analysis is done from the variation of bond lengths, bond angles, and octahedral tilt. Fig-\ref{fig:N14}(b) and (c) depicts the normalized stretching and angular distortions of SRCO15 (SrRuO$_3$), with respect to the value at 300 K (250 K), as a function of $T/T_C$. In the vicinity of FM ordering the relative stretching distortion ($\Delta$/$\Delta_{300K}$) shows a peak for both SRO and SRCO15. Such peak in $\Delta$ neat $T_C$ is expected. Below 100 K in SRCO15, $\Delta$ shows an increasing trend whereas SrRuO$_3$ exhibits a peak-like feature (in the locality of CFM). While the relative angular distortion ($\Sigma$/$\Sigma_{300K}$) show an increasing trend down to FM ordering, and then it freezes down to $\sim$ 150 K. Lowering the temperature results decrease in $\Sigma$/$\Sigma_{300K}$ down to 100 K and then again freezes. Fig-\ref{fig:N14}(d) displays the relative inter-octahedral tilt ($\theta$/$\theta_{300 K}$) against the $T/T_C$. The octahedral tilt is calculated from the Ru-(O1/O2)-Ru bond angles as: tilt = $\frac{180 - \angle{(Ru-(O1/O2)-Ru)}}{2}$ (angle $\theta$ schematically shown in Fig-\ref{fig:N1}(d)). With lowering the temperature, the octahedral tilt along the apical plane initially increases by 2.5\% and gets frozen in the vicinity of $T_C$ followed by a reduction (while along the basal plane, it is increased) of 2.5\% down to 100 K. Further lowering temperature neither affects the apical nor basal tilt. Such freezing in the angular distortion and octahedral tilt indicates the formation of an ordered phase (in this case it is FM). To get more insight into magnetotransport study on various SRCO samples is carried out up to magnetic field 5 T. 
\begin{figure*}
\begin{center}
\includegraphics[width=0.9\linewidth]{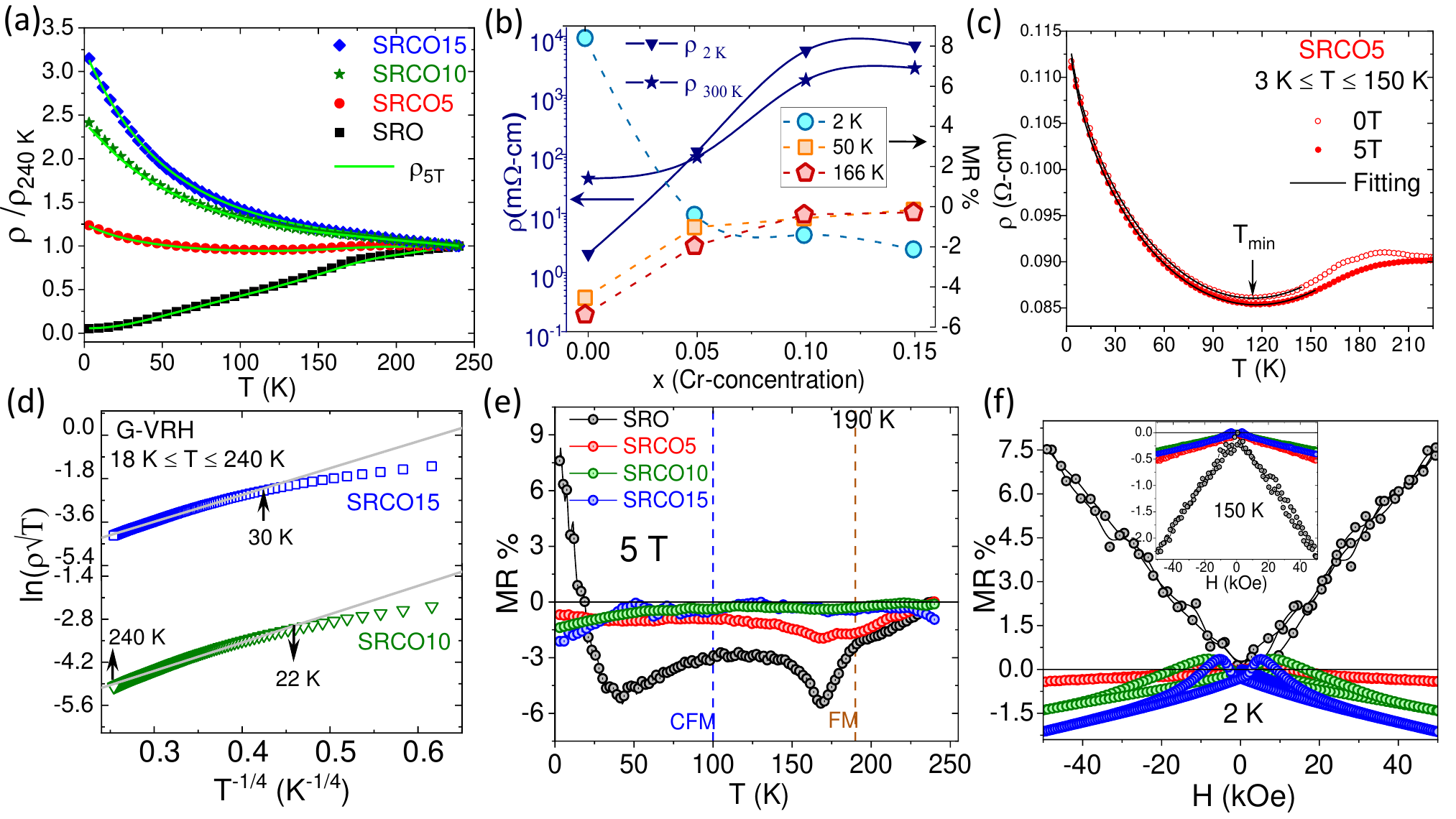} 
 \caption{(color online) (a) The temperature-dependent normalised electrical resistivity ($\rho/\rho_{250 K}$) of SRCO samples at 0 T (coloured data points) and 5 T (solid green line). (b) Variation of $\rho_{2K}$, $\rho_{300 K}$ (left y-axis), and MR\% at 2 K, 50 K, and 166 K (right y-axis) with increasing Cr concentrations. (c) Resistivity fitting of SRCO5 sample with WL-model. (d) G-VRH models fitting of resistivity in SRCO10 and SRCO15. (e) $MR$ ($\frac{\rho(5T)-\rho(0)}{\rho(0)}$ $\times$100)\% vs temperature data; vertical dotted lines represent the onset of CFM and FM. (f) MR\% vs magnetic field at 2 K; inset shows MR in the vicinity of FM ordering for SRCO samples.}
\label{fig:RT}
\end{center}
\end{figure*}
\subsection{Transport measurements-} 
Figure-\ref{fig:RT}(a) displays the temperature-dependent normalized resistivity ($\rho/\rho_{250}$ K). The parent compound SRO exhibits the metallic behaviour down to 2 K (with residual resistivity ratio (RRR) $\sim$ 20, with a slope change at $T_C$ (166 K). A gradual increase in resistivity with Cr-doping indicates the evolution of non-metallic behaviour. Fig-\ref{fig:RT}(b) depicts the variation of $\rho_{300K}$ and $\rho_{2K}$ with respect to Cr doping. The two and four orders of magnitude increase in the $\rho_{300K}$ and $\rho_{2K}$ respectively, is estimated at 15\% Cr doping in SRO. The increase in resistivity as well as the low-temperature resistivity upturn arises due to the disorder and electron correlation effect induced by Cr doping. The role of disorder and electron correlation on electrical transport can be understood by appropriate fitting of the resistivity data with various models. The parent compound shows FL to Non-FL behaviour $\sim$ 45 K (fitting not shown) as reported earlier \cite{kostic1998non,sow2012structural}. A resistivity minima is noticed at $T_{min}$ $\sim$ 117 K for x=0.05. In an itinerant electron system with such low Cr doping induces weak localization (WL) \cite{lee1985disordered}. On the other hand, electron-electron interaction (EEI) in such system also drives low-temperature resistivity upturn \cite{dai1992electrical}. In correlated metals with low disorder, WL \cite{lee1985disordered} and EEI \cite{dai1992electrical} drive the insulating state at low temperatures. In such case the resistivity upturn can be fitted using the following relation \cite{sarkar2015correlation}:
\begin{equation}
\rho = \frac{1}{1/\rho_0 +aT^{p/2}+bT^{1/2}}+cT^2.
\label{eq:3}
\end{equation}
Here $\rho_0$ refers to the residual resistivity, the term $aT^{p/2}$ and $bT^{1/2}$ represents WL and EEI effect respectively. The value of $a$ and $b$ gives an idea about the strength of WL and EEI respectively. In the WL term, $p$ depends on the scattering rate of the electron’s dephasing mechanism and dimensionality of the system \cite{dai1992electrical}. The value of $p$ indicates the nature of WL to be electron-electron (p=2) or electron-phonon (p=3) driven. The $cT^2$ term in Eq-\ref{eq:3} represents EEI scattering at high-temperatures. 
Fig-\ref{fig:RT}(c) shows the resistivity fitting of SRCO5 using Eq-\ref{eq:3}. The fitting parameters of resistivity at 0 and 5 T are tabulated in Table-\ref{t:6}. The value of $p$ turns out to be 2.98. Similar value of $p$ has been reported for Ba and Ti doped SrRuO$_3$ as well \cite{sarkar2015correlation}, indicates that the WL is mainly driven by the electron-phonon interaction. However, it is evident from the value of $b$ ($\sim$ 200$a$, see Table-\ref{t:6}) that EEI dominates in the overall conduction process. For x $>$ 0.05 the electron transport follows the modified Mott-Variable Range Hopping also known as Greaves-VRH (G-VRH) model \cite{greaves1973small,lee2002electron}, where the resistivity can be expressed as 
\begin{equation}
\rho(T) = A\sqrt{T} \exp \Big[\Big(\frac{T_G}{T}\Big)^{1/4}\Big]
\label{eq:4}
\end{equation}
where $T_G$ is Greaves characteristic temperature and $A$ is pre factor. Fig-\ref{fig:RT}(d) shows the G-VRH model fitting in a high-temperature regime for SRCO10 ($T_G$ = 8.7$\times$10$^{3}$ K) and SRCO15 ($T_G$ = 1.7$\times$10$^{4}$ K). The characteristic temperature $T_G$ is inversely proportional to the localization length of conduction electron \cite{lee2002electron,sarkar2015correlation}. With Cr-doping the increased value of $T_G$ indicates that the localization length of conduction electron decreases, which suggests the system is getting localised with increasing the Cr-doping.\par Further to study the magnetic contribution in transport the magnetoresistance measurement is performed. Fig-\ref{fig:RT}(e) depicts the temperature dependent magnetoresistance ($MR\%$ = $\frac{\rho(H)-\rho(0)}{\rho(0)}$ $\times$100), at 5 T in SRCO. Two negative peaks in MR data are observed for the parent compound SRO. The first peak arises owing to the spin fluctuation from PM to FM phase transition at $\sim$ 166 K. The second peak $\sim$ 50 K is attributed to the CFM. With small Cr-doping (x = 0.05), the MR peak at FM order is shifted to a higher temperature ($\sim$ 190 K), and a second peak at 50 K is suppressed. However, both of these negative peaks almost disappear in the case of SRCO10 and SRCO15. Interestingly for SRO a crossover from negative MR to positive MR is noticed near 10 K. Such crossover in MR behavior is also reflected from the R-H data measured at various temperatures as shown in Fig-\ref{fig:RT}(f). For SRO a large non-saturating positive linear MR is observed at 2K. However Cr-doped SRCO samples exhibit a negative MR. The negative MR for ferromagnetic metallic systems can be explained by using spin fluctuation theory. While the origin of positive MR in bulk SrRuO$_3$ is not known. Although, it can be attributed to CFM in SRO. It has to be noted that, the 5 T field is not sufficient to align the complex magnetic structure, which can effectively increase the spin disorder scattering and hence the positive MR at low temperatures \cite{nagasawa1972magnetoresistance,yamada1973magnetoresistance}. However the positive MR in SrRuO$_3$ thin films has quantum origin such as weak antilocalisation due to large spin-orbit coupling, orbital effect, and Weyl fermionic nature \cite{gausepohl1995magnetoresistance, gunnarsson2012anisotropic,wang2023symmetry}. The hysteretic nature in MR is observed for all SRCO samples. This hysteretic nature increases with increasing Cr doping. The value of MR is reduced to $\sim$ -2\% for SRCO15 at 2 K as shown in Fig-\ref{fig:RT}(b) (right y-axis). At 150 K the hysteresis in MR decreases and a weak negative MR ($\sim$ -0.5\%) is noticed in the case of SRCO samples (inset Fig-\ref{fig:RT}(f)). Whereas large negative MR (-2\%) is observed for the parent compound (corresponds to an FM ordering at $\sim$ 166 K). The temperature-dependent MR response of SRCO is nicely correlated with the temperature-dependent structural evolution obtained from ND data. In a nutshell, Cr doping: (i) distorts the crystal structure, (ii) dilutes the total number of magnetic ions and (iii) increases the disorder in SRO. These three facts effectively install a non-metallic state with reduced ferromagnetic strength. This study indicates that the structural modulation with temperature enhances the T$_C$.  
\begin{table}
\centering
 \caption{Fitting parameters of SRCO5 resistivity at 0 and 5T.}
\begin{tabular}{|c|c|c|c|c|c|}\hline
 &$\rho_0$ & $a$ ($\Omega$cm$^{-1}$ & $p$ & $b$ ($\Omega$cm$^{-1}$&$c$ ($\Omega$cm\\
 &($\Omega$cm) & K$^{-p/2}$) & & K$^{-1/2}$)&K$^{-2}$)\\
 \hline
0T& 0.121&0.0015&2.98&0.32&8.9$\times$ $10^{-7}$\\
 5T&0.119&0.0014&2.92&0.34&7.9$\times$ $10^{-7}$\\
 \hline
 \end{tabular}
 \label{t:6}
\end{table} 
\section{CONCLUSION-} Here we observe enhanced Curie-temperature by 22 K in Cr doped SrRuO$_3$ and installs non-metallic behaviour as the delocalised Ru-4d orbitals are replaced by localised Cr-3d. The structural analysis in SrRu$_{1-x}$Cr$_x$O$_3$ (0 $<$ x $<$ 0.15) finds orthorhombic ($Pnma$) crystal structures with squeezed volume. The stretching distortions of the RuO$_6$ octahedra increase up to five times at x = 0.15. XPS core level spectra find mixed valance state of Cr (as Cr$^{3+}$ and Cr$^{6+}$) and the valence band spectra exhibit spectral weight shifting of delocalised electron towards the localised state. The low-temperature M(T) fit indicates a larger $J$ value for SRCO than SRO supporting higher Curie temperature. ND refinements of SRCO15 find that the magnetic moment is aligned along the $b$-axis with $M_s$ = 1.3 $\mu_B$/Ru$^{4+}$ which closely matches with the experimental value of the bulk magnetic moment at 2 K, 6 T (1.1 $\mu_B$/f.u.). Temperature-dependent ND analysis reveals a structural modulation which is closely associated with the enhanced $T_C$. The unit cell volume minima of SRCO15 shifts by 22 K compared to SRO, exactly mimics the $T_C$ shift. In transport, the resistivity upturn for x = 0.05 and the non-metallic behaviour for x $>$ 0.05 is fitted in terms of disorder and localization picture. The magnetotransport data reveals a gradual suppression of spin scattering near FM as well as CFM ordering and an evolution of a non-metallic state with Cr-doping. Thus, the itinerant-localised duality of the d-electron becomes crucial in deciding the fate of the electronic as well as magnetic character. 

\section*{ACKNOWLEDGMENTS-} The authors acknowledge SRG SERB Grants (SRG-2019-001104, CRG-2022-005726, EEQ-2022-000883), India, and Initiation Grant (IITK-2019-037), IIT Kanpur, for financial support. Special Thanks to S. Yonezawa (Kyoto University, Japan) for the magnetic and magnetotransport measurements. 

\bibliography{bib3}
\end{document}